\documentstyle[epsfig,twocolumn,aps,amssymb]{revtex}
\begin{document}
\draft

\title{Finite-temperature simulations of the scissors mode in Bose-Einstein 
 condensed gases}
\author{B. Jackson and E. Zaremba}
\address{Department of Physics, Queen's University, Kingston, Ontario 
 K7L 3N6, Canada.}
\date{\today}
\maketitle
\begin{abstract}

The dynamics of a trapped Bose-condensed gas at finite temperatures is
described by a generalized Gross-Pitaevskii equation for the condensate
order parameter and a semi-classical kinetic equation for the thermal
cloud, solved using $N$-body simulations. The two components are
coupled by mean fields as well as collisional processes that transfer
atoms between the two. We use this scheme to investigate scissors
modes in anisotropic traps as a function of temperature. Frequency 
shifts and damping rates of the condensate mode are extracted, and are 
found to be in good agreement with recent experiments.

\end{abstract}

The experimental observation of the scissors mode in a gaseous Bose-Einstein 
condensate (BEC) \cite{marago00} provided a characteristic signature of  
superfluidity in this system~\cite{guery-odelin99}. 
The scissors mode is excited by an angular displacement of the gas  
relative to an anisotropic confining potential. Above the BEC 
transition temperature, $T_c$, the resulting oscillation generally 
consists of the superposition of two modes, a high-frequency mode which
reduces to an irrotational quadrupole mode in the limit of an isotropic
trap, and a low-frequency mode which in the same limit corresponds to a
pure rotation at zero frequency~\cite{guery-odelin99}.
In contrast, a pure condensate exhibits only one oscillation frequency,
indicating the irrotational, and therefore 
superfluid, nature of the gas. An important issue in superfluid systems 
is the transition between these regimes with increasing temperature. 
This question has recently been addressed experimentally for trapped gases 
\cite{marago01}, stimulating the need for a consistent theoretical 
description of the observed behavior.
     
It is well established that many properties of the condensate at very low
temperatures can be described by the Gross-Pitaevskii (GP) equation 
\cite{dalfovo99}, which is a nonlinear Schr\"{o}dinger equation for the 
condensate wavefunction, $\Phi ({\mathbf{r}},t)$. The equation treats the 
condensate as a classical field and neglects quantum and thermal 
fluctuations. Consequently, the theory breaks down at higher temperatures
($T>0.5 T_c$) where the non-condensed component of the cloud is significant.    
Including the thermal component in a consistent manner is a considerable
challenge. Most calculations, such as those based
on the Hartree-Fock-Bogoliubov (HFB) equations \cite{hutchinson97,dodd98},
fail to capture the full collective dynamics of the thermal cloud,
particularly its back-action on the condensate.

An approach which allows one to treat the dynamics of both components
simultaneously was developed previously~\cite{zaremba99}. The resulting
equations of motion reduce to a generalized GP equation for the 
condensate 
\begin{eqnarray}
 i\hbar \frac{\partial}{\partial t} \Phi ({\mathbf{r}},t) = \Bigg(
 -\frac{\hbar^2}{2m} \nabla^2 + V ({\mathbf{r}}) + g
 [n_c ({\mathbf{r}},t) + \nonumber \\
 2 \tilde{n} ({\mathbf{r}},t)]-iR ({\mathbf{r}},t) \Bigg) 
 \Phi({\mathbf{r}},t),
\label{eq:GP-gendamp}
\end{eqnarray}
and a semiclassical 
Boltzmann kinetic equation for the thermal cloud 
\begin{equation} 
 \frac{\partial f}{\partial t} + \frac{{\mathbf{p}}}{m} \cdot \nabla f
 - \nabla U_{\rm eff} \cdot \nabla_{\mathbf{p}} f 
 = C_{12} [f] + C_{22} [f].
\label{eq:kinetic}
\end{equation}
$n_c({\mathbf{r}},t)= |\Phi({\mathbf{r}},t)|^2$ and $\tilde{n} 
({\mathbf{r}},t)$ are the condensate and non-condensate densities,
respectively, and the mean field acting on the thermal atoms is given by
$U_{\rm eff} = V + 2g[n_c+\tilde n]$. Apart from mean field effects,
Eqs.~(\ref{eq:GP-gendamp}) and (\ref{eq:kinetic}) are linked by
the non-Hermitian source term, $R ({\mathbf{r}},t)$, which accounts 
for the transfer of
atoms between the condensate and thermal cloud and is defined in terms
of the $C_{12}[f]$ collision integral by
\begin{equation}
 R ({\mathbf{r}},t) = \frac{\hbar}{2n_c} \int \frac{d{\mathbf{p}}}
 {(2\pi\hbar)^3} C_{12} [f]\,.
\label{eq:diss-term}
\end{equation}
In arriving at this set of equations, several approximations have been
made. The use of a contact interatomic interaction, $g \delta 
({\mathbf{r}}-{\mathbf{r}}')$ ($g=4\pi \hbar^2 a / m$,
where $a$ is the atomic $s$-wave scattering length), is standard. More
importantly, the thermal excitations are treated in the Hartree-Fock
approximation. Furthermore, we invoke the Popov approximation, where 
the `anomalous' density 
$\tilde {m}$ is neglected. 
This is in part motivated by physical concerns: inclusion of $\tilde{m}$
leads to unphysical low-momentum gaps in the energy spectrum, as well as
infra-red and ultra-violet divergences. These inconsistencies can be
removed either perturbatively~\cite{morgan00}, or through the use of
refined kinetic equations~\cite{walser99}.
The latter may provide a possible future extension of our work.   

In this paper we apply the above set of equations to the calculation of
the scissors mode in an anisotropic harmonic trap [$V({\bf r}) =
m(\omega_\rho^2 \rho^2 + \omega_z^2 z^2)/2$] as a function of 
temperature for experimentally 
relevant parameters. The GP equation is solved using a FFT 
split-operator method~\cite{fft}, whereas the kinetic equation is solved
by performing
a classical simulation~\cite{wu97,bird94} in which the thermal 
phase-space density $f({\bf p},{\bf r},t)$ is represented by 
an $N$-body system of discrete particles. Collisions between atoms are
naturally treated in this representation by Monte-Carlo computation of
the $C_{22}$ and $C_{12}$ collision integrals. The complete dynamical 
problem becomes computationally feasible within this scheme, and has the
additional advantage of providing an intuitive physical picture for the
thermal gas dynamics. The damping rates and frequency shifts that we
calculate for the scissors mode are in good agreement with experiment,
indicating that our approach is valid over a wide range of temperatures.
The simulations also yield the quadrupole response function of the 
system, which clearly illustrates the transition between superfluid 
and rigid-body behavior with increasing temperature. 

Our simulations involve propagation over a sequence of time steps,
$\Delta t$. The trajectories of the thermal atoms are obtained by
solving Newton's equations of motion with forces defined by the
mean field potential $U_{\rm eff}$. To ensure energy conservation 
the velocities and positions of the atoms are updated at each time-step
using a second-order simplectic integrator \cite{yoshida93}. To define
the mean field of a thermal cloud consisting of discrete particles,
we first allocate atoms to 
grid-points ${\mathbf{r}}_{jkl}$ using a cloud-in-cell approach 
\cite{hockney81}, and then convolve this density distribution with a
Gaussian function $G({\mathbf{r}})\sim e^{-r^2/\eta^2}$. This
effectively is a smoothening procedure, which
is equivalent to assuming that interactions involving thermal atoms 
have a finite range $\eta$. For consistency, the $n_c$ term appearing 
in $U_{\rm eff}$ is also convolved.

The Monte-Carlo computation of the $C_{22}$ collision term, describing
collisions between two non-condensed atoms that both remain in the 
thermal cloud, was discussed in an earlier paper~\cite{jackson01}. 
However, the $C_{12}$ collision term, which transfers atoms between 
the two 
components, was neglected. We now include this term, but as the method 
is somewhat involved we will detail it elsewhere \cite{jackson_la}. 
Briefly, one can write (\ref{eq:diss-term}) as $R=R^{\rm in}-
R^{\rm out}$, where $R^{\rm in}$ represents collisions between two
thermal particles that lead to absorption of one by the condensate, 
while $R^{\rm out}$ refers to the inverse process. In terms 
of our Monte-Carlo simulations, one approximates the integrals by a sum
over particles around a grid-point: $R ({\mathbf{r}}_{jkl}, t) = 
(\hbar/2n_c) \sum_i \Delta P_i$, where 
$\Delta P_i=P_i^{\rm in} - P_i^{\rm out}$. The term $P_i^{\rm in}$
($P_i^{\rm out}$) represents the probability that a given thermal atom with
velocity ${\mathbf{v}}_i$ will collide during the time step, leading to
a net transfer of particles in (out) of the condensate. Both probabilities
are proportional to $n_c$ and the Bose collision cross-section, 
$\sigma=8\pi a^2$. They also depend upon the local condensate velocity 
${\mathbf{v}}_c$ and the phase-space densities of the final thermal 
states, randomly selected to satisfy momentum and energy conservation.
  
In summary, the simulations consist of the following sequence.
In a given time step, the thermal particle phase-space
coordinates are first updated as discussed above. The 
probabilities of $C_{12}$ collisions are next calculated for each atom.
A random number $R \in [0,1]$
is chosen, and if $|\Delta P_i|>R$, atoms are added ($\Delta P_i<0$) or 
removed ($\Delta P_i>0$) appropriately. The quantity $\Delta P_i$ is 
also accumulated to define the dissipative term in 
(\ref{eq:GP-gendamp}). We then treat the $C_{22}$ collisions 
\cite{jackson01}. Finally, the GP equation is propagated; the
dissipative term, $R$, leads to a continuous change in normalization of the 
wavefunction which is consistent with the discrete addition or removal 
of particles from the thermal cloud.

Experimentally \cite{marago00,marago01} the condensate is produced in a
disk-shaped anisotropic trap ($\omega_z \sim \sqrt{8} \omega_\rho$), 
which 
is adiabatically tilted to make an angle $\theta_0$ with respect to its
original orientation. The scissors mode is then excited by suddenly 
switching the trap to an angle $-\theta_0$, so that the condensate and 
thermal cloud oscillate about this new equilibrium position. We 
simulate this scenario by first finding the equilibrium condensate 
and non-condensate density profiles for a particular temperature $T$ 
using a self-consistent semiclassical procedure \cite{jackson01}. 
A sample of test particles is then chosen to simulate the
phase space distribution,
where to minimize statistical fluctuations, ten times the physical 
number of thermal atoms is used.  Initial particle positions and 
momenta are randomly selected by a rejection method from a Bose 
distribution $f({\mathbf{p}},{\mathbf{r}},t)=[z^{-1} \exp(\beta p^2/2m)
-1]^ {-1}$, where $\beta \equiv 1/k_B T$ and $z({\mathbf{r}})=
\exp ( -\beta \{
U_{\rm eff} ({\mathbf{r}}) - \mu \})$ is the position-dependent fugacity
($\mu$ is the condensate chemical potential). The particle coordinates,
as well as the condensate density, are then rotated through an angle 
$2\theta_0$ about the $y$ axis relative to the trap potential 
$V({\mathbf{r}})$.

Starting with these initial conditions, quadrupole moments can be 
calculated separately for the condensate ($Q_c(t)=\int d{\bf r}\,xz
n_c$) and thermal cloud (${\tilde{Q}}(t)=\sum_{i=1}^{\tilde{N}} 
x_i z_i$). To make contact with experiment we define a rotation angle 
for each component, $\theta_\alpha(t) = Q_\alpha(t)/Q_\alpha^0$, where 
$Q_\alpha^0 = \langle x^2 - z^2 \rangle^0_\alpha$ is the
equilibrium quadrupole moment of the $\alpha$-th
component~\cite{zambelli01}. For a pure condensate ($T=0$) consisting 
of $N=2\times 10^4$ atoms, the quadrupole moment is found to oscillate
with almost constant amplitude (where small fluctuations arise from 
weak excitation of other condensate modes) at a single frequency of 
$\omega_{sc}=2.9886 \omega_\rho$, which is about $1.5 \%$ larger than 
the Thomas-Fermi result 
$\omega_{sc}=(\omega_\rho^2 + \omega_z^2)^{1/2}$~\cite{guery-odelin99} 
due to finite 
number effects. Above $T_c$ the Bose gas oscillation exhibits two 
frequencies with approximately equal amplitudes. Our simulations yield 
frequencies that are very close to 
those found experimentally and predicted analytically: 
$\omega_\pm = |\omega_\rho \pm \omega_z|$.
In addition, the thermal cloud oscillation is weakly damped by $C_{22}$ 
collisions, over a timescale which is similar for both modes and is of
the order of several collision times.    

\begin{figure}
\centering
\psfig{file=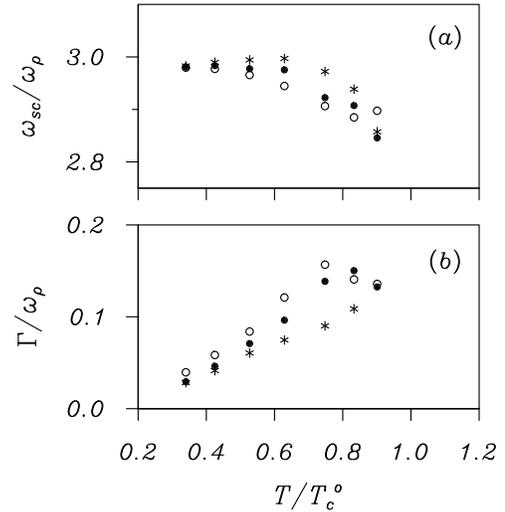, scale=0.35, angle=0, bbllx=50, bblly=115, 
 bburx=570, bbury=665}
\caption{\label{dampfreq-fnT1}
 Frequency (a) and damping rate (b) of the condensate scissors mode
 as a function of temperature for $N=5 \times 10^4$ total particles. Note
 that the temperature is normalized by the transition temperature of the
 ideal gas, $T_c^0$. Each plot shows the effect of inclusion of each term in
 the kinetic equation (\ref{eq:kinetic}): the free-streaming operator alone
 ($\ast$); the free-streaming and $C_{22}$ terms ($\bullet$); and all terms 
 ($\circ$).}
\end{figure}

Below $T_c$, our simulations describe the dynamics of both components.
For most temperatures we find that the condensate and thermal cloud
modes are essentially excited independently, indicating that the two 
components are only weakly coupled. Nevertheless, the condensate 
oscillation experiences
significant damping from interactions with the thermal component.
To quantify the damping rate and frequency of the condensate mode, 
the data is 
fit to a single exponentially-decaying sinusoidal function in order to
make contact with the experimental analysis~\cite{marago01}.
Fig.~\ref{dampfreq-fnT1} shows
results for the condensate mode as a function of temperature, for a
fixed total number of atoms, $N=5 \times 10^4$. To separate the effects
of each term in the Boltzmann equation (\ref{eq:kinetic}) on the 
dynamics, simulations are performed with no collisions, with only
$C_{22}$ collisions, and with both $C_{12}$ and $C_{22}$. At low
temperatures ($T<0.6 T_c^0$) we see that the damping is predominately 
due to collisionless Landau damping, where mean field
interactions between the condensate and thermal atoms
transfer energy from the collective mode to single particle 
excitations. 
When $C_{22}$ collisions are included, an increase in 
damping is observed. These collisions can only affect the condensate 
indirectly through their equilibrating effect on the nonequilibrium
distribution of thermal atoms.
The $C_{12}$ term leads to additional damping at low $T$ by further 
promoting the equilibration of the condensate and thermal cloud. This 
source of damping is comparable in magnitude to that of the $C_{22}$ 
collisions but
is small compared to Landau damping, in agreement with a previous study
of collective modes under the assumption of a static thermal 
cloud \cite{williams01}. The effect of $C_{12}$ collisions can also be
inferred by monitoring the number of condensate atoms as a function of 
time. Since
the initial angular displacement of the cloud increases its total
energy, the gas must equilibrate to a higher final temperature with a
corresponding reduction in the condensate number. This is seen in the 
simulations. 

At higher temperatures ($T > 0.6 T_c^0$), collisional effects increase 
in importance, but the situation becomes more complicated as $T_c^0$ 
is approached. The more massive thermal cloud
begins to drive the condensate at its own scissors mode frequencies, and
as a result, a single damped sinusoid is a poor fit to the theoretical
data in this regime. Effectively, the condensate oscillations are 
sustained over longer times and the damping appears to saturate.

There are also observable effects on the condensate mode frequency. In
the collisionless limit, we see from Fig.~1(a) that the frequency at
first increases with increasing temperature as a result of the
decreasing number of condensate atoms (i.e., the frequency shifts away 
from
the TF limit). This trend is reversed by collisions, as well as 
at higher temperatures where the
thermal cloud becomes significant. Since
the condensate tends to drag part of the thermal cloud along with it,
its effective inertia is increased. One would therefore expect a
lowering of its normal mode frequency, as observed.

\begin{figure}
\centering
\psfig{file=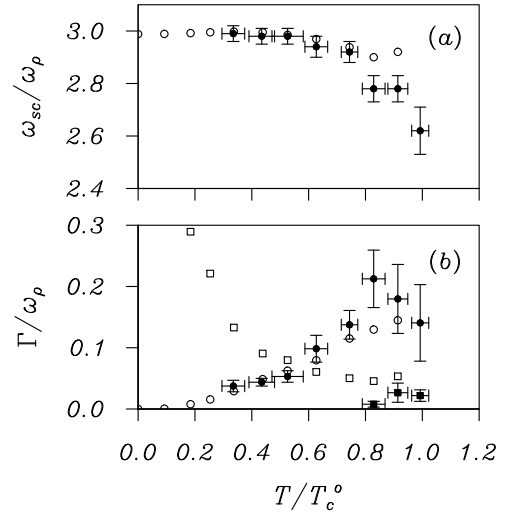, scale=0.35, angle=0, bbllx=50, bblly=115, 
 bburx=570, bbury=665}
\caption{\label{dampfreq_fnT2} 
 Frequency (a) and damping rate (b) of the scissors modes for a variable
 total number of atoms, intended to simulate the experimental 
 parameters. The condensate mode is indicated by open (theory) and solid
 (experiment) circles. The open squares in (b) show the calculated 
average 
 damping rate of the two thermal cloud modes, while the solid squares 
 are the corresponding experimental values.}
\end{figure}

Use of evaporative cooling in the actual experiments \cite{marago01} 
meant that the total number of atoms varied with temperature from
$N \simeq 2\times 10^4$ at low $T$ to $N \simeq 10^5$ close to $T_c^0$. 
Since a small but significant dependence on number is found in our full
simulations, we have taken this variation into account in our comparison
with experiment. Our results in Fig.~\ref{dampfreq_fnT2} are in very 
good agreement with experiment for $T<0.8 T_c^0$ but deviate from
experiment in both frequency and damping at the three highest
temperatures. However the experimental error bars are
particularly large for these points, which may reflect difficulties in
extracting values of the damping rate and frequency from fits to the
data over a limited timescale. We note that three damped
sinusoids are needed to fit the theoretical data in this regime, as the
condensate is strongly coupled to the thermal cloud. In addition, we do
not see a condensate at the highest temperature point ($T\simeq T_c^0$),
which may indicate systematic errors in the experimental temperature
scale or possibly limitations of the semiclassical approximation.

\begin{figure}
\centering
\psfig{file=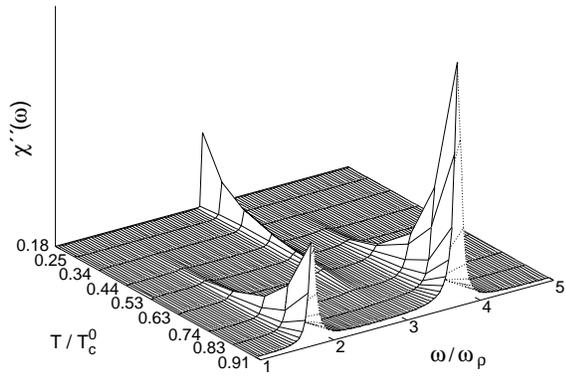, scale=0.7, angle=0, bbllx=70, bblly=60, 
 bburx=375, bbury=270}
\caption{\label{quadresp}
 Quadrupole response function $\chi'' (\omega)$ for the system, as a function
 of frequency and temperature. The parameters match those in Fig.\
 \ref{dampfreq_fnT2}.}
\end{figure}

A quantity of considerable interest is the quadrupole response function
$\chi''(\omega)$, which physically describes the energy absorption of each
mode under a harmonic perturbation. Zambelli and Stringari \cite{zambelli01}
demonstrated that this can be related to the moment of intertia of the system,
$\Theta$, by the expression
\begin{equation}
\frac{\Theta}{\Theta_{\rm rigid}} = (\omega_z^2 - \omega_\rho^2)^2 
\frac{\int {\rm d}\omega \, \chi''(\omega) / \omega^3}{\int 
{\rm d}\omega \, \chi'' (\omega) \, \omega}, 
\label{eq:moi}
\end{equation}
where $\Theta_{\rm rigid}$ is the moment of inertia of the corresponding
rigid body. For a sudden rotation of the trap, $\chi'' (\omega) \propto
\omega \, \Re \{ Q (\omega) \}$, where $Q (\omega)$ is the Fourier
transform of the time-dependent quadrupole moment. We extract this 
quantity by fitting three damped sinusoids to the calculated total 
quadrupole moment. Fig.~\ref{quadresp} shows results as a function of 
temperature. At lower temperatures, the condensate mode is dominant and
$\chi''(\omega)$ exhibits a single peak. For very low $T$ the thermal 
cloud is in fact strongly coupled to the condensate, and oscillates at 
the condensate frequency with a small phase shift that accounts for 
condensate damping. However, as the temperature is raised, the strength
of the condensate mode diminishes, and the spectral density is dominated
by the two thermal cloud modes at $\omega_{\pm}$, as would be expected 
for $\Theta \simeq \Theta_{\rm rigid}$~\cite{zambelli01}. 
In contrast, the small $T$ behavior is consistent with a 
superfluid moment of intertia, $\Theta_{\rm sf}
< \Theta_{\rm rigid}$. In principle, one could calculate the moment of
inertia over the entire temperature range using (\ref{eq:moi}).
However, the damped nature of the modes leads to Lorentzian spectral
densities for which the required frequency moments are undefined.
Nevertheless, one can see qualitatively that $\Theta_{\rm sf} < 
\Theta < \Theta_{\rm rigid}$ in the intermediate region.

To summarize, we have simulated the scissors modes in a 
finite-temperature Bose gas using a coupled FFT/Monte-Carlo scheme. 
Our approach takes into account fully the dynamical mean fields acting
between the condensate and thermal cloud as well as all collisional
processes that are physically relevant. We find very good agreement 
with experiment over a wide range of temperatures, with only minor
discrepancies near to $T_c^0$ where both theory and experiment are more
difficult to analyze quantitatively. These same methods can be used to
study other collective modes, as well as problems such as
condensate growth.

We acknowledge financial support from NSERC of Canada, and
the use of the HPCVL computing facilities at Queen's. We thank A. 
Griffin, J. Williams
and T. Nikuni for valuable discussions, and O. Marag\`o and C. Foot for
providing experimental data. BJ would also like to thank C. Adams and
J. McCann for their help and support over an extended period.

   
\begin{references}
\bibitem{fft} Energy conservation and significant gains in stability 
 are realized using a second-order approximation for the evolution
 operator: $\Phi (t+\Delta t)=e^{-i \tilde{H} \Delta t} \Phi (t)$, where
 $\tilde{H} (t) = [3H(t)-H(t-\Delta t)]/2$. Typically $\omega_\rho
 \Delta t = 1.5 \times 10^{-3}$, while a 3D spatial grid of $64\times 
 64 \times 32$ points is utilized.
\end{references}
\end{document}